\documentclass[conference]{IEEEtran}
\IEEEoverridecommandlockouts

\usepackage{cite}
\usepackage{amsmath,amssymb,amsfonts}
\usepackage{algorithm}
\usepackage[noend]{algpseudocode}
\usepackage{graphicx}
\usepackage{booktabs}
\usepackage{textcomp}
\usepackage{xcolor}

\algrenewcommand\algorithmicrequire{\textbf{Input:}}
\algrenewcommand\algorithmicensure{\textbf{Output:}}

\begin{document}

\title{Federated Multi-Agent Soft Actor--Critic for IRS-Assisted Cognitive Radio Networks under Partial Observability}

\author{%
\IEEEauthorblockN{Satvika Pullisani\IEEEauthorrefmark{1},
Usama Mir\IEEEauthorrefmark{1}, \textit{Senior Member, IEEE}, and
Ubaid Abbasi\IEEEauthorrefmark{2}}
\IEEEauthorblockA{\IEEEauthorrefmark{1}Department of Computer Science and Physics, Wilfrid Laurier University, Brantford, ON, Canada}
\IEEEauthorblockA{\IEEEauthorrefmark{2}Department of Science, Northwestern Polytechnic, Grande Prairie, AB, Canada\\
Corresponding author: Usama Mir (e-mail: umir@wlu.ca)}
}

\maketitle

\begin{abstract}
Intelligent reflecting surfaces (IRSs) can improve spectrum efficiency and interference control in cognitive radio networks (CRNs) by reconfiguring the propagation environment. This paper presents FMASAC, a federated multi-agent soft actor--critic framework for IRS-assisted CRNs under partial observability, in which distributed secondary users (SUs) jointly select their transmit power and the phase shifts of an assigned subset of IRS elements using only local observations. Primary user (PU) protection is incorporated through an interference-penalized reward, and SUs coordinate by periodically averaging model parameters at a federated server, without exchanging raw observations, channel state information, or replay data. We formulate the problem as a partially observable Markov decision process, specify the local learning and aggregation procedure, and characterize its per-round communication cost. Illustrative performance trends contrast FMASAC with centralized SAC, independent SAC, IRS-fixed multi-agent reinforcement learning, and a heuristic CRN strategy. We also outline the simulation-based evaluation and the open challenges, namely constraint handling, agent heterogeneity, and scalability, that are needed to validate the framework.
\end{abstract}


\section{Introduction}

Intelligent reflecting surfaces (IRSs) have emerged as a promising technology for improving wireless communication through programmable control of the propagation environment~\cite{taha2020}. By tuning passive reflecting elements, an IRS can strengthen desired links and suppress harmful interference without active radio frequency (RF) transmission, making it a low-power enabler for spectrum-efficient wireless networks. However, IRS performance depends on selecting phase-shift configurations under dynamic channel conditions, which becomes difficult when channel state information (CSI) is incomplete or rapidly changing. This limitation motivates learning-based control methods such as deep reinforcement learning (DRL), which can adapt IRS decisions through interaction with the wireless environment.

DRL has been widely investigated for IRS optimization because it can learn reflection strategies through environment interaction rather than explicit channel models~\cite{omid2022}. DRL-based and limited-feedback IRS configurations have shown strong potential in smart radio environments, but they focus on standalone IRS control and do not jointly model multiple secondary users (SUs), primary user (PU) protection, and decentralized learning~\cite{wang2022,zhu2024}. This motivates the use of IRS and DRL in cognitive radio networks (CRNs), where licensed PUs own the spectrum and unlicensed SUs opportunistically access it without causing harmful interference to the PUs. How efficiently licensed spectrum is used also depends on how operators price and share access to it~\cite{mir2013}. In this setting, IRS-assisted DRL can improve SU communication by adapting reflection decisions to changing wireless conditions. At the same time, the problem becomes more challenging because multiple SUs must make transmission decisions from limited local observations while still protecting the PUs.

IRS-assisted CRNs have attracted attention because an IRS can enhance SU links while reducing interference toward the PUs. DRL-based resource allocation and adaptive spectrum access improve IRS-assisted spectrum utilization and adaptability to changing channel and spectrum conditions, but they rely on centralized learning and lack federated multi-agent coordination across distributed SUs~\cite{guo2022,srivastava2025}. IRS-assisted spectrum sensing and primary--secondary transmission improve sensing and transmission performance but do not address decentralized joint learning of SU policies and IRS-aware control~\cite{liang2025}. DRL-based IRS-assisted CRN optimization shows the value of dynamically adjusting IRS configurations to the wireless environment~\cite{zhong2022}, and hierarchical DRL has been applied to secure resource allocation in multi-IRS spectrum-sharing networks~\cite{wang2024}; however, these approaches remain single-agent or centralized and assume broader state availability than distributed CRNs provide.

When multiple SUs operate simultaneously in shared spectrum, the action of one SU affects its own link quality, the interference experienced by other SUs, and PU protection. This interaction makes decentralized decision-making difficult under partial observability. Cooperative multi-agent architectures, in which SU agents negotiate spectrum use among themselves without a central controller, were an early answer to this problem~\cite{mir2010comas}. Multi-agent reinforcement learning (MARL) addresses this problem by allowing multiple agents to learn transmission policies while interacting through a shared wireless environment, and multi-agent DRL has already been used to adapt IRS configurations~\cite{monga2025}. Cooperative MARL has improved spectrum sensing efficiency, dynamic spectrum access, and multi-user access decisions~\cite{gao2021,tan2022,naparstek2019,vangaru2025}, but existing MARL methods often rely on centralized training, shared observations, or frequent information exchange, which increases communication overhead and limits practicality in distributed CRNs. This limitation motivates federated learning, where SUs coordinate by sharing model parameters instead of raw observations or local interaction data.

Federated learning also supports privacy-preserving collaboration. Prior studies on IRS-assisted cognitive radio communication~\cite{nguyen2025} mainly focus on communication-level analysis and performance modeling, and IRS-assisted federated learning studies improve communication efficiency and secure parameter aggregation~\cite{wu2023,mao2025}. However, these works do not address decentralized multi-agent learning, spectrum-sharing dynamics, or IRS-assisted reinforcement learning under partial observability. Conventional federated learning alone also does not model the interaction among multiple agents competing in shared spectrum. Federated multi-agent actor--critic learning has been applied to dynamic spectrum sharing~\cite{yang2023}, and federated multi-agent DRL has been surveyed for wireless communications~\cite{oliveira2024}, but neither provides a unified IRS-assisted CRN framework with soft actor--critic (SAC)-based continuous control.

To the best of our knowledge, no existing study jointly addresses IRS-assisted CRN optimization, federated learning, multi-agent SAC (MASAC)-based continuous control, and partial observability in one framework. To address this gap, this paper proposes a federated MASAC (FMASAC) framework in which each SU trains locally while a federated server aggregates model parameters without raw data exchange. The main contributions are as follows:
\begin{itemize}
\item We propose FMASAC, a federated MASAC framework for IRS-assisted CRNs that operates without raw data sharing or global CSI.
\item We give a joint formulation of SU transmit power and IRS phase control under a PU interference constraint and partial observability, including an explicit IRS element-partitioning model that resolves how multiple agents act on a single IRS.
\item We model the problem as a partially observable Markov decision process (POMDP), specify the reward, the SAC updates, and the federated aggregation rule, and characterize the per-round communication cost.
\item We present illustrative performance trends against four benchmarks and outline the simulation-based evaluation and open challenges required to validate the framework.
\end{itemize}

The remainder of this paper is organized as follows. Section~\ref{sec:system} presents the system model and POMDP formulation, Section~\ref{sec:framework} describes FMASAC, Section~\ref{sec:trends} presents illustrative performance trends, Section~\ref{sec:limitations} discusses limitations and the evaluation plan, and Section~\ref{sec:conclusion} concludes the paper.

\section{System Model and Problem Formulation}\label{sec:system}

We consider an IRS-assisted CRN with one primary base station (PBS) serving one PU, and $M$ distributed SU transmitter--receiver pairs that opportunistically access the licensed spectrum while keeping the aggregate interference at the PU below a prescribed threshold. An IRS with $N$ passive reflecting elements assists the secondary transmissions. The principal symbols are summarized in Table~\ref{tab:notation}.

\begin{table}[t]
\caption{Main Notation}
\label{tab:notation}
\centering
\footnotesize
\begin{tabular}{@{}ll@{}}
\toprule
\multicolumn{2}{@{}l}{\textbf{Network and IRS parameters}}\\
$M$, $N$ & Number of SUs; number of IRS elements\\
$\boldsymbol{\Theta}(t)$, $\theta_n(t)$ & IRS reflection matrix; phase of element $n$\\
$\mathcal{N}_i$ & Subset of IRS elements controlled by SU $i$\\
$p_i(t)$, $P_{\max}$ & Transmit power of SU $i$; power budget\\
$\mathrm{SINR}_i(t)$, $R_i(t)$ & SINR and rate at the receiver of SU $i$\\
$I_i(t)$, $I_{\max}$ & Interference from SU $i$ at the PU; threshold\\
\midrule
\multicolumn{2}{@{}l}{\textbf{POMDP and learning variables}}\\
$s(t)$, $o_i(t)$ & Global state; local observation of SU $i$\\
$a_i(t)$, $\mathbf{a}(t)$ & Action of SU $i$; joint action\\
$r_i(t)$, $\lambda$ & Reward of SU $i$; interference penalty weight\\
$\gamma$, $\xi(t)$ & Discount factor; channel variation and noise\\
\midrule
\multicolumn{2}{@{}l}{\textbf{SAC and federated learning parameters}}\\
$\pi_{\varphi_i}$, $\varphi_i$ & Actor policy of SU $i$ and its parameters\\
$Q_{\psi_i^{(k)}}$, $k\in\{1,2\}$ & Twin critics of SU $i$ and their parameters\\
$\mathcal{D}_i$ & Local replay buffer of SU $i$\\
$\alpha$ & Entropy coefficient\\
$K$, $R$ & Local steps per round; number of rounds\\
$w_i$ & Aggregation weight of SU $i$\\
\bottomrule
\end{tabular}
\end{table}

\subsection{Channel and IRS Model}

\begin{figure}[t]
\centering
\includegraphics[width=0.95\columnwidth]{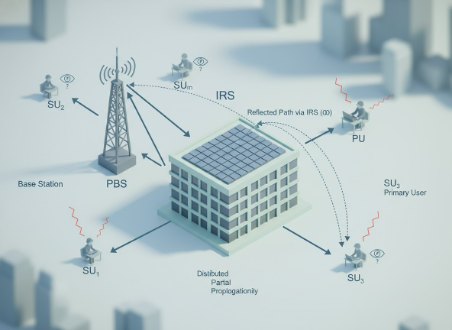}
\caption{IRS-assisted CRN topology with direct, reflected, and interference paths.}
\label{fig:topology}
\end{figure}

As shown in Fig.~\ref{fig:topology}, the IRS creates a controllable reflected path that can strengthen useful SU signals while reducing interference toward the PU. All channels follow Rayleigh small-scale fading with log-distance path loss. The IRS reflection matrix is
\begin{equation}
\boldsymbol{\Theta}(t) = \mathrm{diag}\left(e^{j\theta_1(t)}, \ldots, e^{j\theta_N(t)}\right),
\label{eq:irs}
\end{equation}
where $\theta_n(t)\in[0,2\pi)$ is the phase shift of the $n$th element. The effective channel from the transmitter of SU $j$ to the receiver of SU $i$ is $h_{ji}(t) = h^{\mathrm{d}}_{ji}(t) + \mathbf{g}_i^{H}(t)\boldsymbol{\Theta}(t)\mathbf{f}_j(t)$, where $h^{\mathrm{d}}_{ji}$ is the direct link, $\mathbf{f}_j$ is the transmitter-to-IRS channel, and $\mathbf{g}_i$ is the IRS-to-receiver channel. The channel $h_{j\mathrm{P}}(t)$ to the PU is defined in the same way. The SINR at the receiver of SU $i$ is
\begin{equation}
\mathrm{SINR}_i(t) = \frac{p_i |h_{ii}|^2}{\sum_{j\neq i} p_j |h_{ji}|^2 + p_0 |h_{0i}|^2 + \sigma^2},
\label{eq:sinr}
\end{equation}
where $p_0$ and $h_{0i}$ are the PBS transmit power and its channel to SU receiver $i$, and $\sigma^2$ is the noise power. The achievable rate is $R_i(t) = \log_2\!\left(1 + \mathrm{SINR}_i(t)\right)$, and the interference caused by SU $i$ at the PU is $I_i(t) = p_i(t)|h_{i\mathrm{P}}(t)|^2$.

Because a single physical IRS cannot apply conflicting phase requests from several agents, the $N$ elements are partitioned into disjoint subsets $\mathcal{N}_1,\ldots,\mathcal{N}_M$, and SU $i$ controls only the phases of the elements in $\mathcal{N}_i$. This keeps each agent's action space fixed and avoids a central IRS arbitration step; alternatives such as a dedicated IRS controller agent are discussed in Section~\ref{sec:limitations}.

\subsection{POMDP Formulation}

Each SU has access only to local information, namely received signal strength, partial CSI of its own links, a PU interference indicator broadcast by the PBS, and limited IRS feedback. The global state and the other SUs' actions and observations are unavailable. The problem is therefore modeled as a POMDP $\langle \mathcal{S}, \mathcal{A}, \mathcal{O}, P, \mathcal{R}, \gamma\rangle$, where $\mathcal{S}$, $\mathcal{A}$, $\mathcal{O}$, $P$, and $\mathcal{R}$ denote the global state space, joint action space, observation space, transition probability, and reward function, and $\gamma\in(0,1)$ is the discount factor. At time $t$, SU $i$ observes $o_i(t)\in\mathcal{O}$ and selects
\begin{equation}
a_i(t) = \left[p_i(t),\, \{\theta_n(t)\}_{n\in\mathcal{N}_i}\right],
\label{eq:action}
\end{equation}
with $p_i(t)\in[0, P_{\max}]$. The joint action is $\mathbf{a}(t) = (a_1(t),\ldots,a_M(t))$, and the global state evolves as
\begin{equation}
s(t+1) = f\left(s(t), \mathbf{a}(t)\right) + \xi(t),
\label{eq:dynamics}
\end{equation}
where $f(\cdot)$ captures the environment dynamics and $\xi(t)$ captures channel variations and noise. The aggregate PU interference must satisfy
\begin{equation}
\sum_{i=1}^{M} I_i(t) \le I_{\max}.
\label{eq:constraint}
\end{equation}

Constraint~\eqref{eq:constraint} is enforced through the reward. Each SU receives
\begin{equation}
r_i(t) = R_i(t) - \lambda \left[\sum_{j=1}^{M} I_j(t) - I_{\max}\right]^{+},
\label{eq:reward}
\end{equation}
where $[x]^{+} = \max(0,x)$ and $\lambda>0$ weights PU protection against SU rate. The violation term is available to every SU through the PBS interference indicator, so no SU needs the other SUs' channels or actions. The penalty enforces~\eqref{eq:constraint} softly: violations are discouraged but not ruled out during learning. The objective is to learn policies $\{\pi_i\}$ that maximize the expected discounted reward
\begin{equation}
\max_{\{\pi_i\}}\; \mathbb{E}\left[\sum_{t=0}^{\infty} \gamma^{t} r_i(t)\right].
\label{eq:objective}
\end{equation}
In the federated setting, each SU learns from its own interaction data and periodically uploads only model parameters, and the server returns an aggregated model.

\section{Proposed Federated MASAC Framework}\label{sec:framework}

\begin{figure}[t]
\centering
\includegraphics[width=0.95\columnwidth]{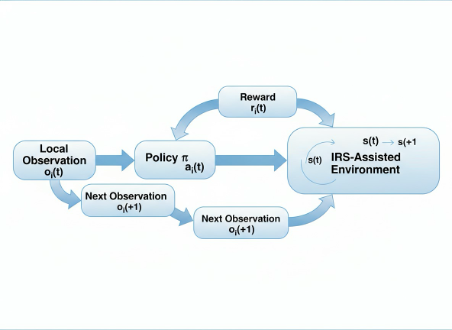}
\caption{POMDP decision loop with federated aggregation.}
\label{fig:loop}
\end{figure}

Building on the POMDP formulation, FMASAC combines local continuous-control learning at each SU with periodic federated aggregation, enabling distributed policy improvement without exchanging raw data. As shown in Fig.~\ref{fig:loop}, each SU interacts with the IRS-assisted environment using only local observations and updates its policy and value networks from a local replay buffer. A federated server periodically aggregates the local model parameters and broadcasts the result, preserving distributed operation while improving cross-agent consistency.

\subsection{Local SAC Learning}

Each SU runs a local SAC agent with actor $\pi_{\varphi_i}$ and twin critics $Q_{\psi_i^{(1)}}$ and $Q_{\psi_i^{(2)}}$, both conditioned on local observations and actions only. SAC suits this problem because transmit power and phase shifts are continuous decisions, and entropy regularization improves exploration and stability under partial observability. The critic target is
\begin{equation}
y = r + \gamma\left(\min_{k} Q_{\bar{\psi}_i^{(k)}}(o', \tilde{a}') - \alpha \log \pi_{\varphi_i}(\tilde{a}'\,|\,o')\right),
\label{eq:target}
\end{equation}
where $\tilde{a}'\sim\pi_{\varphi_i}(\cdot\,|\,o')$ and $\bar{\psi}_i^{(k)}$ are target-critic parameters. Each critic minimizes the soft Bellman loss
\begin{equation}
L(\psi_i^{(k)}) = \mathbb{E}_{(o,a,r,o')\sim\mathcal{D}_i}\left[\left(Q_{\psi_i^{(k)}}(o,a) - y\right)^{2}\right],
\label{eq:critic}
\end{equation}
and the actor maximizes
\begin{equation}
J(\varphi_i) = \mathbb{E}_{\substack{o\sim\mathcal{D}_i\\ a\sim\pi_{\varphi_i}}}\left[\min_{k} Q_{\psi_i^{(k)}}(o,a) - \alpha\log\pi_{\varphi_i}(a\,|\,o)\right],
\label{eq:actor}
\end{equation}
where $\alpha$ controls the exploration--exploitation trade-off.

\subsection{Federated Aggregation}

After $K$ local steps, each SU uploads its actor and critic parameters, and the server forms the weighted average
\begin{equation}
\bar{\varphi} = \sum_{i=1}^{M} w_i\, \varphi_i, \qquad w_i = \frac{|\mathcal{D}_i|}{\sum_{j=1}^{M} |\mathcal{D}_j|},
\label{eq:fedavg}
\end{equation}
with the same rule applied to each critic. The aggregated parameters are broadcast to all SUs. Averaging assumes that SUs are sufficiently similar for a shared model to be useful; we normalize observations per SU to support this, and Section~\ref{sec:limitations} discusses heterogeneous settings where personalized aggregation is more appropriate. Algorithm~\ref{alg:fmasac} summarizes the procedure.

\begin{algorithm}[t]
\caption{FMASAC Training}
\label{alg:fmasac}
\small
\begin{algorithmic}[1]
\Require $M$ SUs, local steps $K$, rounds $R$, weights $\{w_i\}$
\State Initialize $\varphi_i$, $\psi_i^{(1)}$, $\psi_i^{(2)}$, target critics, and $\mathcal{D}_i$ for each SU $i$
\For{round $r = 1,\ldots,R$}
  \For{each SU $i$ in parallel}
    \For{$k = 1,\ldots,K$}
      \State Observe $o_i(t)$; sample $a_i(t)\sim\pi_{\varphi_i}(\cdot\,|\,o_i(t))$
      \State Execute $a_i(t)$; receive $r_i(t)$ and $o_i(t+1)$
      \State Store $(o_i(t), a_i(t), r_i(t), o_i(t+1))$ in $\mathcal{D}_i$
      \State Update critics via \eqref{eq:critic} and actor via \eqref{eq:actor}
      \State Soft-update target critics
    \EndFor
    \State Upload $\{\varphi_i, \psi_i^{(1)}, \psi_i^{(2)}\}$ to the server
  \EndFor
  \State Server aggregates parameters using \eqref{eq:fedavg}
  \State Server broadcasts aggregated parameters to all SUs
\EndFor
\Ensure Aggregated actor and critic parameters
\end{algorithmic}
\end{algorithm}

\subsection{Communication Cost}\label{sec:comm}

Let $|\varphi|$ and $|\psi|$ denote the numbers of actor and per-critic parameters. In each round, every SU uploads and downloads $|\varphi| + 2|\psi|$ parameters, so the total traffic per round is $2M(|\varphi| + 2|\psi|)$, and $R$ rounds cost $2MR(|\varphi| + 2|\psi|)$. The cost is independent of the replay-buffer size and of the number of interaction steps $K$, and no CSI or observations leave the SU. Increasing $K$ reduces the number of rounds needed for a given amount of local experience but lets local models drift further apart between aggregations, which is the trade-off examined in the planned evaluation.

\section{Illustrative Performance Trends}\label{sec:trends}

\subsection{Scope of This Section}

The curves in Figs.~\ref{fig:throughput}--\ref{fig:variance} were produced in Python (NumPy, Matplotlib) from controlled synthetic performance traces that encode the qualitative behavior expected of each method. They are not the outcome of training agents in a simulated network and should be read as illustrations of the trade-offs that FMASAC targets, not as measured performance. A simulation-based evaluation is in progress, and its design is described in Section~\ref{sec:limitations}.

The illustrated setting consists of one PBS, one PU, and distributed SUs over a 1~km$^2$ area, with a fixed-position IRS and Rayleigh fading with log-distance path loss, over 300 training episodes. FMASAC is contrasted with four benchmarks: centralized SAC (a single agent with global information access, which serves as a reference but violates the distributed assumption), independent SAC (local learning without aggregation), IRS-fixed MARL (learned power control with a fixed IRS configuration), and a heuristic CRN strategy (conservative transmission under PU protection). Four metrics are considered: average SU throughput (spectrum utilization), PU interference level normalized to $I_{\max}$ (PU protection), average cumulative reward (convergence), and action variance across SUs (policy consistency).

\subsection{Illustrated Trends}

\begin{figure}[t]
\centering
\includegraphics[width=\columnwidth]{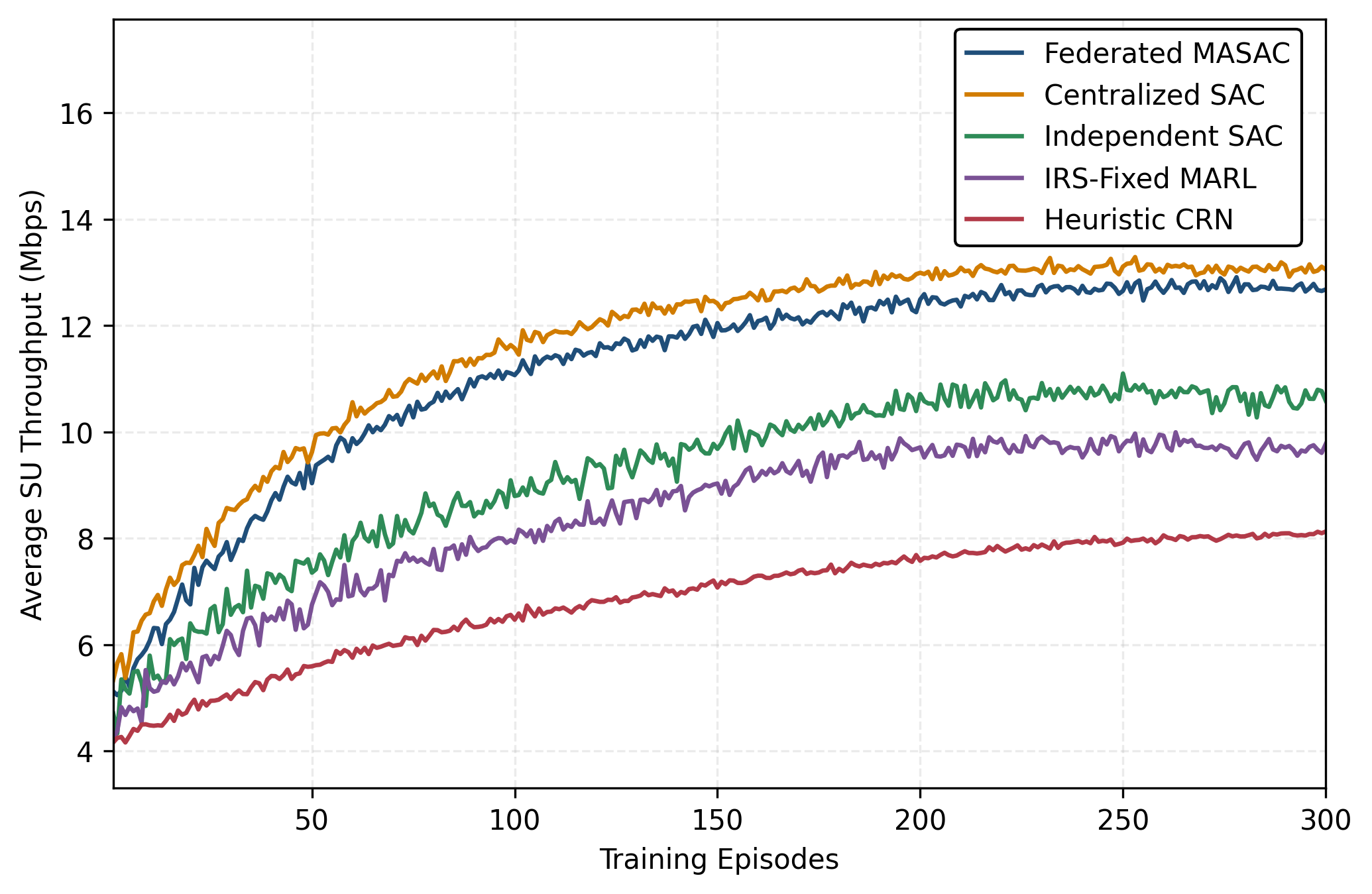}
\caption{Average SU throughput versus training episodes (illustrative).}
\label{fig:throughput}
\end{figure}

\begin{figure}[t]
\centering
\includegraphics[width=\columnwidth]{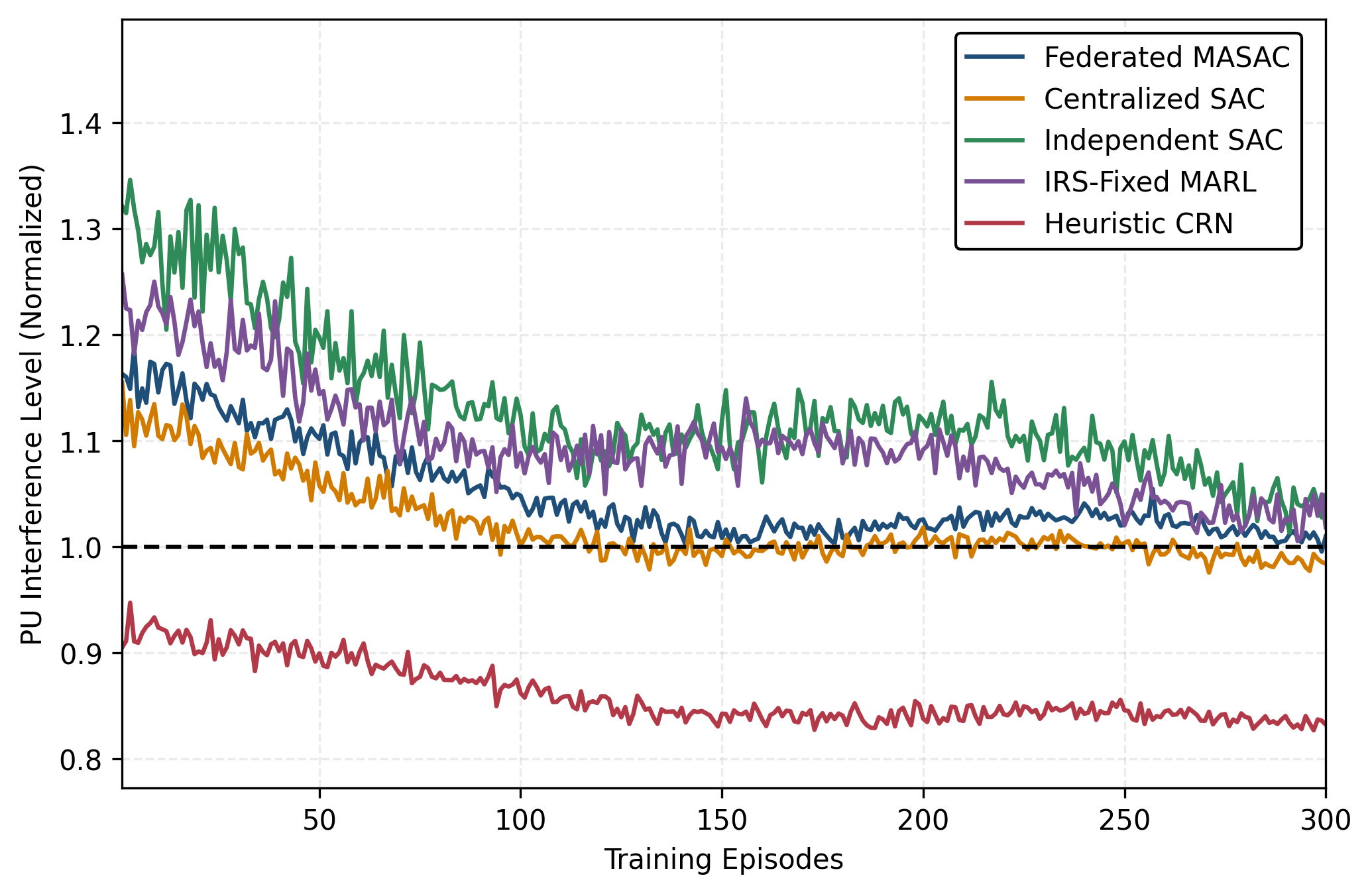}
\caption{PU interference level, normalized to $I_{\max}$, versus training episodes (illustrative).}
\label{fig:interference}
\end{figure}

Fig.~\ref{fig:throughput} illustrates the expected throughput ordering. FMASAC is shown above independent SAC, IRS-fixed MARL, and the heuristic strategy, reflecting the intended benefit of combining IRS-aware control with federated aggregation, while centralized SAC acts as an upper reference because it uses information unavailable to distributed SUs.

Fig.~\ref{fig:interference} illustrates PU interference. The learning-based methods start above the normalized threshold and approach it as training proceeds, with FMASAC shown fluctuating less than independent SAC and IRS-fixed MARL. Only the heuristic strategy stays below the threshold throughout, at the cost of lower throughput. This pattern reflects the soft, penalty-based enforcement of~\eqref{eq:constraint} in~\eqref{eq:reward}: early in training, constraint violations are expected. Applications that require guaranteed protection would need the constrained formulations discussed in Section~\ref{sec:limitations}.

\begin{figure}[t]
\centering
\includegraphics[width=\columnwidth]{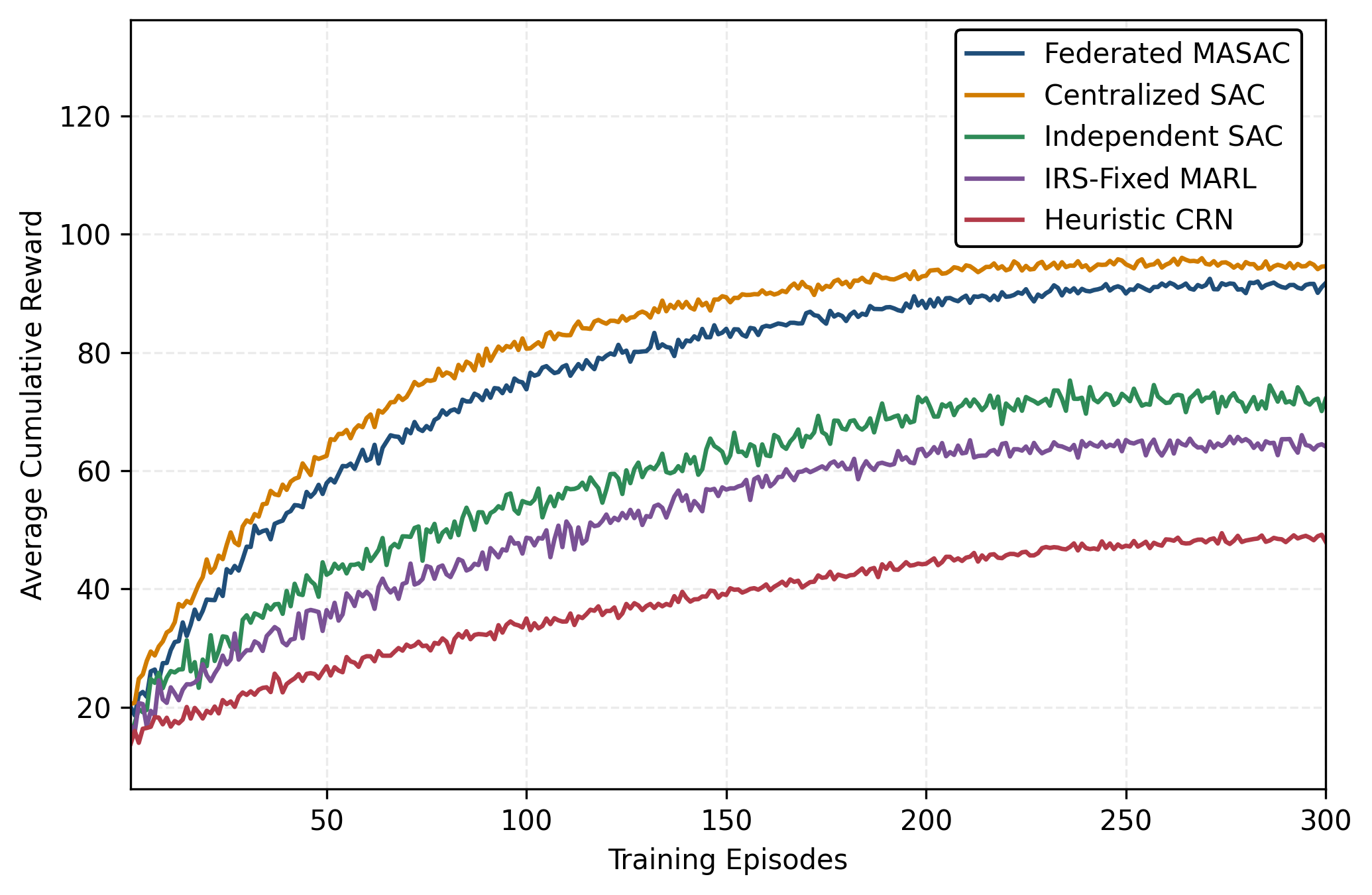}
\caption{Average cumulative reward versus training episodes (illustrative).}
\label{fig:reward}
\end{figure}

\begin{figure}[t]
\centering
\includegraphics[width=\columnwidth]{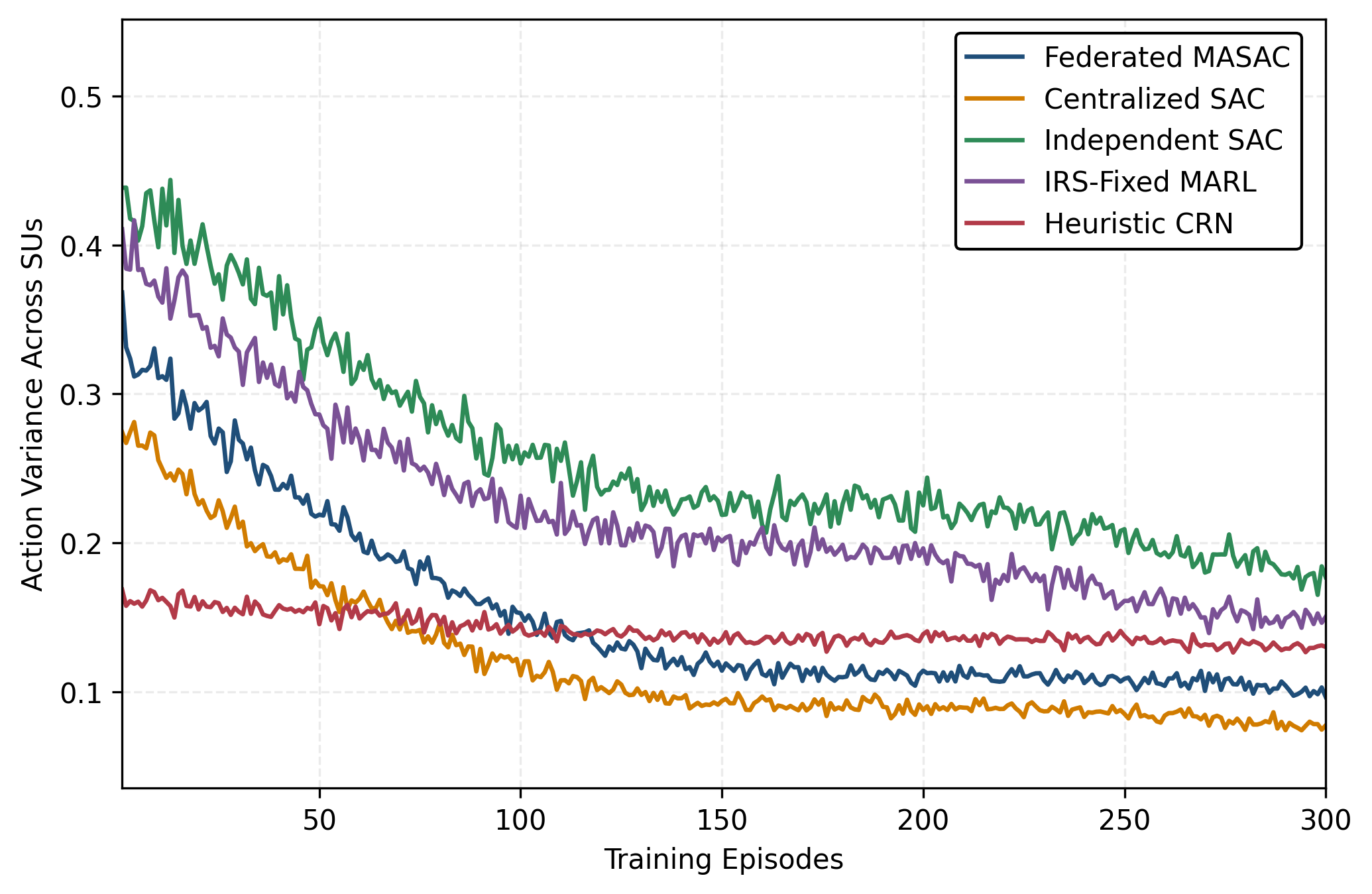}
\caption{Action variance across SUs versus training episodes (illustrative).}
\label{fig:variance}
\end{figure}

Fig.~\ref{fig:reward} illustrates convergence. Periodic aggregation is expected to smooth learning in the non-stationary multi-SU environment, so FMASAC is shown converging more smoothly than the decentralized baselines, while centralized SAC converges well but relies on global information.

Fig.~\ref{fig:variance} illustrates policy consistency through action variance across SUs. Aggregation is expected to reduce this variance relative to uncoordinated learning, particularly independent SAC, which would translate into more consistent multi-user behavior and less interference variation. Whether these trends hold in practice, and by how much, is precisely what the simulation study must establish.

\section{Limitations and Evaluation Plan}\label{sec:limitations}

The framework leaves several questions open, which define the planned evaluation.
\begin{itemize}
\item \textbf{Simulation-based validation.} Implement the channel model of Section~\ref{sec:system} with SAC agents and report, for stated values of $M$, $N$, $K$, $P_{\max}$, $\sigma^2$, $\lambda$, node positions, and network hyperparameters, the mean and confidence intervals over independent random seeds.
\item \textbf{Aggregation interval and overhead.} Vary $K$ and measure throughput and constraint satisfaction against the communication cost derived in Section~\ref{sec:comm}.
\item \textbf{Scalability.} Vary $M$ and $N$ to test whether the element-partitioning model and parameter averaging remain effective as the network grows.
\item \textbf{Constraint handling.} Compare the penalty in~\eqref{eq:reward} with primal--dual or Lagrangian SAC, and report the probability of violating~\eqref{eq:constraint} rather than only the average interference level.
\item \textbf{Heterogeneity.} Evaluate non-identically distributed SU locations and channels, and compare plain averaging in~\eqref{eq:fedavg} with personalized federated aggregation.
\item \textbf{IRS control architecture.} Compare element partitioning with a dedicated IRS controller agent that sets all $N$ phases from aggregated SU feedback.
\end{itemize}

\section{Conclusion}\label{sec:conclusion}

This paper studied spectrum access and IRS-assisted control in CRNs under partial observability, where distributed SUs must learn from local observations while protecting the PU. We proposed FMASAC, in which each SU performs local soft actor--critic learning over its transmit power and an assigned subset of IRS elements, and a federated server periodically averages model parameters without raw data exchange. We formulated the problem as a POMDP with an interference-penalized reward, specified the learning and aggregation procedure, and characterized its communication cost. Illustrative trends indicate the trade-offs the framework is designed to improve: throughput, PU protection, convergence smoothness, and multi-user consistency relative to decentralized baselines. Validating these trends through simulation, studying the trade-off between aggregation frequency, communication overhead, and learning performance, and strengthening constraint guarantees are the subjects of ongoing work.


\end{document}